\documentclass[12pt]{iopart}
\newcommand{\beq}{\begin{equation}}
\newcommand{\eeq}{\end{equation}}
\newcommand{\md}{{\rm d}}
\newcommand{\imu}{{\rm i}}
\newcommand{\mrm}[1]{\mathrm{#1}}
\newcommand{\mcal}[1]{\mathcal{#1}}

\begin{document}
\title[The Hilbert-Space Structure of Non-Hermitian Theories with Real
Spectra]{The Hilbert-Space Structure of Non-Hermitian Theories with 
Real Spectra\footnote{Talk given by R. Kretschmer at the {\em 1st
International Workshop on Pseudo-Hermitian Hamiltonians in Quantum 
Physics}, Prague, Czech Republic, June 16-17, 2003.}}
\author{Ralph Kretschmer{\rm \dag\S} and 
Lech Szymanowski$\|${\rm \P}}
\address{\dag\ Fachbereich Physik, Universit\"at Siegen, Germany}
\address{$\|$ Soltan Institute for Nuclear Studies, Warsaw, Poland}
\eads{\S\ \mailto{kretschm@hepth2.physik.uni-siegen.de},\\
\mailto{\P\ lech.szymanowski@fuw.edu.pl}}

\begin{abstract}
We investigate the quantum-mechanical interpretation of models with
non-Hermitian Hamiltonians and real spectra. After describing a 
general framework to reformulate such models in terms of Hermitian 
Hamiltonians defined on the Hilbert space $L_2(- \infty, \infty)$, we 
discuss the significance of the algebra of physical observables. 
\end{abstract}


\section{Introduction}
\label{s1}

In the past few years, models with non-Hermitian Hamiltonians (the 
term non-Hermiticity meant here in the sense of the space 
$L_2(- \infty, \infty)$ of square-integrable functions) have attracted 
a lot of interest, because many examples are known in which such 
models have real spectra \cite{Bender1, Cannata, Znojil}. Therefore, 
they may describe realistic physical systems. 

Despite this, the physical interpretation of these models remains 
unclear: The eigenstates $\psi_n$ of a non-Hermitian Hamiltonian $H$ 
are not mutually orthogonal (their scalar product may even be 
undefined) and the time evolution generated by $H$ is non-unitary, so 
that the usual probabilistic interpretation of wave functions cannot 
be applied. In addition, in some models \cite{Bender1, Cannata} it is 
necessary to extend the definition of position-space wave functions to 
complex values of the coordinate. This means that the wave functions 
are not elements of the Hilbert space $L_2(- \infty, \infty)$, so that 
non-Hermiticity in the sense of the space $L_2$ has no obvious meaning
here.

Recently, it seems that a consensus has been reached that---provided
the spectrum of the Hamiltonian is real---one can always define a
positive-definite scalar product under which the eigenstates are
orthogonal. Among the approaches investigated so far is
\begin{itemize}
\item[$\bullet$] the notion of pseudo-Hermiticity, advocated in the
work of Mostafazadeh \cite{Mostafazadeh} (see also \cite{Scholtz} for 
an early discussion of this concept). Here a metric operator $\eta$ is 
used to define a modified scalar product,
\beq
(\psi, \varphi) := (\psi, \eta \varphi)_{L_2}
\eeq
($(.,.)_{L_2}$ is the scalar product in $L_2$),
\item[$\bullet$] the introduction of a $\mcal{CPT}$ transformation
\cite{Bender2}
\beq
(\psi, \varphi) := \int_C \md x \, [\mcal{CPT} \psi(x)] \varphi(x) 
\quad,
\eeq
and
\item[$\bullet$] the direct construction of the Hilbert space 
$\mcal{H} = \overline{\mrm{span}\{\psi_1, \psi_2, \ldots\}}$ (that is 
the closure of the space of finite superpositions of the 
eigenfunctions) with a scalar product defined by \cite{Kretschmer}
\beq\label{1a}
(\psi_n, \psi_m)_\mcal{H} := \delta_{nm} \quad \mbox{for all } n, m
\quad.
\eeq
\end{itemize}
In this contribution we want to explain the last method in some 
detail. We will only treat the case of a discrete, infinite spectrum
of $H$.

\section{The canonical formulation}
\label{s2}

Definition (\ref{1a}) leads to a separable Hilbert space $\mcal{H}$ in 
which the Hamiltonian $H$ is Hermitian (provided its spectrum is 
real), because for two vectors $\varphi = \sum_n a_n \psi_n, 
\psi = \sum_n b_n \psi_n \in \mcal{H}$ that are in the domain of 
definition of $H$ one finds
\beq
(\varphi, H \psi)_\mcal{H}
= \sum_{n, m} a_n^* b^{\vphantom{*}}_m (\psi_n, H \psi_m)_\mcal{H}
= \sum_{n, m} a_n^* b^{\vphantom{*}}_m E_n \delta_{nm} 
= (H \varphi, \psi)_\mcal{H} \quad.
\eeq
Thus $\mcal{H}$ does not only define a consistent probabilistic
structure for the theory defined by $H$, it also guarantees a unitary
time evolution.

Since all separable, infinite-dimensional Hilbert spaces are unitarily 
equivalent, a map $T: \mcal{H} \to L_2$ exists that fulfills
\beq\label{2a}
(\varphi, \psi)_\mcal{H} = (T \varphi, T \psi)_{L_2}
\quad\mbox{for all } \varphi, \psi \in \mcal{H} 
\eeq
\cite{Prugovecki}. With the help of this transformation, the theory 
can equivalently be defined in the space $L_2$ in terms of the 
operator $\hat{H} = T H T^{- 1}: L_2 \to L_2$ and the eigenstates 
$\hat{\psi}_n = T \psi_n \in L_2$. As a consequence of the unitary
equivalence of $\mcal{H}$ and $L_2$, the operator $\hat{H}$ is 
Hermitian in $L_2$: For $\hat{\varphi}$, $\hat{\psi}$ in the domain of
definition of $\hat{H}$ one has
\begin{eqnarray*}
(\hat{\varphi}, \hat{H} \hat{\psi})_{L_2}
& = & (T^{- 1} \hat{\varphi}, T^{- 1} \hat{H} \hat{\psi})_\mcal{H}
= (T^{- 1} \hat{\varphi}, H T^{- 1} \hat{\psi})_\mcal{H}
= (H T^{- 1} \hat{\varphi}, T^{- 1} \hat{\psi})_\mcal{H} \\
& = & (T^{- 1} \hat{H} \hat{\varphi}, T^{- 1} \hat{\psi})_\mcal{H}
= (\hat{H} \hat{\varphi}, \hat{\psi})_{L_2} \quad.
\end{eqnarray*}

This construction is very general, and one can easily find trivial, in
general physically insignificant transformations $T$. An example is 
the linear map $T: \mcal{H} \to L_2$ that fulfills 
$T \psi_n = \psi_n^{(\mrm{ho})}$ for all $n$, where the 
$\psi_n^{(\mrm{ho})}$ are the eigenstates of the harmonic oscillator. 
Then
\[
\hat{H} \psi^{(\mrm{ho})}_n = T H T^{- 1} T \psi_n = E_n T \psi_n
= E_n \psi^{(\mrm{ho})}_n \quad.
\]

We want to stress that in order to find physically acceptable 
transformations $T$, one has to take into account the fact that the 
Hamiltonian has to be a function of physical observables.

Initially, one usually starts with a Hamiltonian that is of the form
\beq
H(x, p) = p^2 + V(x) \quad,
\eeq
in which $p$ and $x$ are the usual position and momentum operators 
that are Hermitian with respect to $L_2$, $V(x)$ is non-Hermitian. In 
the space $\mcal{H}$, $H$ is Hermitian, but $x$ and $p$ will in 
general be non-Hermitian. (Consider, for example, the model 
investigated by Bender el~al.\ \cite{Bender1}, 
$H = p^2 + x^2 (\imu x)^\nu$, $\nu \geq 0$. Here the operators $H$, 
$x$ and $p$ cannot all be simultaneously Hermitian.) This means that 
within the formulation in $\mcal{H}$, the operators $x$ and $p$ can no 
longer be observables. Thus, the physical meaning of the Hamiltonian 
$H(x, p)$ in the space $\mcal{H}$ is quite unclear. 

In order to understand the physical content of $H(x, p)$, one has to 
express it as a function of two operators $x^\mrm{c}$, $p^\mrm{c}$ 
that correspond to the observables of position and momentum,
\beq\label{2b}
H = \tilde{H}(x^\mrm{c}, p^\mrm{c}) \quad.
\eeq
Necessary conditions for the operators $x^\mrm{c}$, $p^\mrm{c}$ are
that
\begin{itemize}
\item[$\bullet$] $x^\mrm{c}$, $p^\mrm{c}$ are Hermitian in $\mcal{H}$,
\item[$\bullet$] they fulfill canonical commutation relations,
$[x^\mrm{c}, p^\mrm{c}] = \imu$.
\end{itemize}
If such operators are found, one may call the representation 
(\ref{2b}) the {\em canonical formulation} of the model. As we will
show, this formulation leads to some insight into the structure of the
transformation $T$.

Let us illustrate this with a very simple example: The Hamiltonian
\beq\label{2c}
H(x, p) = p^2 + {2 \imu \over x} p - {2 \over x^2} + \omega^2 x^2 
\eeq
is non-Hermitian in $L_2$, but (as will become evident below) has a 
real spectrum and square-integrable eigenfunctions. We claim that in 
the space $\mcal{H}$, it is {\em physically equivalent to the harmonic 
oscillator}. 

The reason is that (\ref{2c}) can be written as
\beq
H = x (p^2 + \omega^2 x^2) x^{- 1} =: x \hat{H} x^{- 1} \quad,
\eeq
i.~e.\ the transformation $T$ between $\mcal{H}$ and $L_2$ can be 
chosen to be $T = x^{-1}$. 

In $\mcal{H}$, which is here defined as the image of $L_2$ under
$T^{- 1}$, the scalar product is given by
\beq
(\varphi, \psi)_\mcal{H} = (T \varphi, T \psi)_{L_2}
= \int\limits_{- \infty}^\infty {\md x \over x^2} \, \varphi^*(x)
\psi(x) \quad.
\eeq
One easily finds that with respect to this scalar product, the
relations $x^\dagger = x$ and $p^\dagger = p + 2 \imu / x$ hold, so
that $(p + \imu / x)^\dagger = p + \imu / x$. In fact, (\ref{2c}) can 
be expressed as
\[
H(x, p) = \left( p + {\imu \over x} \right)^2 + \omega^2 x^2 \quad.
\]
Now define $x^\mrm{c} := x$, $p^\mrm{c} := p + \imu / x$. Note that
\beq\label{2d}
x^\mrm{c} = T^{- 1} x T \quad,\quad
p^\mrm{c} = T^{- 1} p T = x p x^{- 1} \quad,
\eeq
so that $[x^\mrm{c}, p^\mrm{c}] = \imu$. Thus $x^\mrm{c}$ and 
$p^\mrm{c}$ fulfill the two necessary conditions given above.

The canonical formulation of the Hamiltonian (\ref{2c}) is therefore
\beq
H = (p^\mrm{c})^2 + \omega^2 (x^\mrm{c})^2 
\equiv \tilde{H}(x^\mrm{c}, p^\mrm{c}) \quad,
\eeq
which makes it evident that we are describing nothing but a harmonic 
oscillator. A summary of the properties of the various operators is 
given in Table~\ref{t1}.

\begin{table}[t]
\caption{Properties of the operators involved in the canonical
formulation.\label{t1}}
\vspace{2mm}
\small
\begin{center}
\begin{tabular}{|c|c|c|}
\hline
operator & in $L_2$ & in $\mcal{H}$\\
\hline\hline
$H$ & non-Hermitian & Hermitian\\
$x$, $p$ & Hermitian &
non-Herm.\ (in general)\\
$\hat{H} = T H T^{- 1}$ & Hermitian & non-Hermitian\\
$x^\mrm{c} = T^{- 1} x T$ & non-Herm.\ (in general)
& Hermitian\\
$p^\mrm{c} = T^{- 1} p T$ & non-Herm.\ (in general)
& Hermitian\\
\hline
\end{tabular}
\vspace{-1mm}
\end{center}
\end{table}

This simple example demonstrates some general features: Given a 
Hermitian Hamiltonian $\hat{H}(x, p)$ that acts in $L_2$, one can make 
a similarity transformation $H = T^{- 1} \hat{H} T$ with an operator
$T$ that is non-unitary if considered as an endomorphism 
$L_2 \to L_2$. Then the spectrum of $H$ remains real, but the
Hermiticity of $H$ is destroyed. If $\hat{H}(x, p)$ is an analytic 
function of $x$ and $p$, then
\beq
H = T^{- 1} \hat{H}(x, p) T = \hat{H}(T^{- 1} x T, T^{- 1} p T)
= \hat{H}(x^\mrm{c}, p^\mrm{c})
\equiv \tilde{H}(x^\mrm{c}, p^\mrm{c}) \quad,
\eeq
i.~e.\ the canonical formulation can be found by substituting 
$x \to x^\mrm{c}$ and $p \to p^\mrm{c}$ in the $L_2$-Hermitian 
Hamiltonian $\hat{H}(x, p)$.

This emphasizes the importance of the canonical formulation: Turning 
the argument around, one starts with a non-Hermitian Hamiltonian 
$H(x, p)$. Once a set of canonical operators $x^\mrm{c}$ and 
$p^\mrm{c}$ is found, and $H$ has been expressed as a function of 
these operators, $H(x, p) = \hat{H}(x^\mrm{c}, p^\mrm{c})$, the model 
can be formulated in the space $L_2$ as an ordinary quantum-mechanical 
problem with the Hermitian Hamiltonian $\hat{H}(x, p)$ that is a 
function of the usual Hermitian position and momentum operators $x$ 
and $p$. Thus the physical meaning of the model is clear. The 
transformation $T$ has to be chosen such that in addition to 
(\ref{2a}) it also fulfills (\ref{2d}).

If, on the other hand, a canonical formulation cannot be found, the 
physical interpretation of $H(x, p)$ is unclear.

\section{Summary and outlook}
\label{s3}

The canonical formulation of the Hamiltonian is meaningful as a
relation between observables. The non-Hermiticity completely
disappears from the model.

The crucial ingredient of this formulation is the correct choice of
the transformation $T: \mcal{H} \to L_2$ (or equivalently the correct
choice of the Hilbert space $\mcal{H}$). Transformations that render a
given non-Hermitian Hamiltonian $H$ with real spectrum Hermitian can
be easily found, but we require in addition that $T$ maps between $x$,
$x^\mrm{c}$ and $p$, $p^\mrm{c}$, resp., via a similarity
transformation.

One immediate question concerns the uniqueness of $T$. The Stone-von
Neumann uniqueness theorem \cite{Prugovecki} states that all 
irreducible representations of two self-adjoint operators $x^\mrm{c}$ 
and $p^\mrm{c}$ that are defined in a separable Hilbert space and 
fulfill canonical commutation relations are unitarily equivalent. 
Finding such a set of operators removes the arbitrariness in $T$ 
completely (up to unitary equivalence). But here the self-adjointness 
(as opposed to Hermiticity) of $x^\mrm{c}$ and $p^\mrm{c}$ is crucial.

In \cite{Kretschmer} we have given explicit expressions
for $T$ for a number of models that have been recently discussed. For
some of these examples, the canonical formulation shows that they are
physically equivalent to well-known quantum-mechanical problems. In 
other models with non-Hermitian Hamiltonians, interesting effects like 
a transition from real to complex eigenvalues occur \cite{Bender1}, or 
the wave functions are analytically continued along complicated paths 
in the complex domain \cite{Cannata}. We believe that in such cases 
the construction of a canonical formulation may lead to interesting 
new insights.

\ack

R.~K. wishes to thank M. Znojil for organizing this very stimulating
workshop. L.~S. is grateful for the warm hospitality extended to him 
at the Ruhr-Universit\"at Bochum. The work of L.~S. is supported in 
part by the German-Polish scientific and technological agreement (WTZ
Deutschland-Polen).


\section*{References}

\begin{thebibliography}{0}
\bibitem{Bender1} C. M. Bender, S. Boettcher, \textit{Phys.\ Rev.\ 
	Lett.}\ \textbf{80} (1998) 5243; C. M. Bender, S. Boettcher, 
	P. Meisinger, \textit{J. Math.\ Phys.}\ \textbf{40} (1999) 
	2201. 
\bibitem{Cannata} F. Cannata, G. Junker, J. Trost, \textit{Phys.\ 
	Lett.}\ A \textbf{246} (1998) 219. 
\bibitem{Znojil} M. Znojil, \textit{Phys.\ Lett.}\ A \textbf{259} 
	(1999) 220.
\bibitem{Mostafazadeh} A. Mostafazadeh, \textit{J. Math.\ Phys.}\
	\textbf{43} (2002) 205; \textit{ibid.}\ \textbf{43}
	(2002) 2814; \textit{ibid.}\ \textbf{43} (2002) 3944.
\bibitem{Scholtz} F. G. Scholtz, H. B. Geyer, F. J. W. Hahne,
	\textit{Ann.\ Phys., NY\/} \textbf{213} (1992) 74.
\bibitem{Bender2} C. M. Bender, D. C. Brody, H. F. Jones,
	\textit{Phys.\ Rev.\ Lett.}\ \textbf{89} (2002) 270401.
\bibitem{Kretschmer} R. Kretschmer, L. Szymanowski, 
	\textit{Preprint\/} quant-ph/0105054 (2001); 
	\textit{Preprint\/} quant-ph/0305123 (2003).
\bibitem{Prugovecki} E. Prugove\v{c}ki, \textit{Quantum Mechanics in 
	Hilbert Space}, Academic Press, New York, 1981.
\end{thebibliography}
\end{document}